\documentclass[review]{elsarticle}

\usepackage[hyphens]{url}
\usepackage{graphicx}
\usepackage{tabularx}
\usepackage{amsmath,amssymb}
\usepackage{mathtools}

\usepackage{todonotes}

\usepackage{tabularx}
\usepackage{booktabs} 

\usepackage{lineno,hyperref}
\hypersetup{
	colorlinks=false,
	pdfborder={0 0 0},
}
\modulolinenumbers[5]

\journal{Arxiv}

\definecolor{eblue}{RGB}{0, 129, 185}
\let\oldcite\cite
\renewcommand{\cite}[1]{ {\color{eblue}(\oldcite{#1})}}

\biboptions{authoryear}

\usepackage{placeins}
\usepackage{color}
\newif\ifincludecomments
\includecommentstrue

\usepackage{cleveref}
\begin{document}

\begin{frontmatter}

\title{Algorithmic Detection and Analysis of Vaccine-Denialist Sentiment Clusters in Social Networks}

\author[dtu]{Bjarke M\o nsted}
\author[dtu]{Sune Lehmann\corref{mycorrespondingauthor}}

\cortext[mycorrespondingauthor]{Corresponding author}
\ead{sljo@dtu.dk}

\address[dtu]{
Technical University of Denmark, Applied Mathematics and Computer Science,
2800 Lyngby, Denmark
}

\begin{abstract}
Vaccination rates are decreasing in many areas of the world, and outbreaks of preventable diseases tend to follow in areas with particular low rates.
Much research has been devoted to improving our understanding of the motivations behind vaccination decisions and the effects of various types of information offered to skeptics, no large-scale study of the structure of online vaccination discourse have been conducted.

Here, we offer an approach to quantitatively study the vaccine discourse in an online system, exemplified by Twitter. We use train a deep neural network to predict tweet vaccine sentiments, surpassing state-of-the-art performance, attaining two-class accuracy of $90.4\%$, and a three-class F1 of $0.762$.
We identify profiles which consistently produce strongly anti- and pro-vaccine content. We find that strongly anti-vaccine profiles primarily post links to Youtube, and commercial sites that make money on selling alternative health products, representing a conflict of interest.
We also visualize the network of repeated mutual interactions of actors in the vaccine discourse and find that it is highly stratified, with an assortativity coefficient of $r = .813$.
\end{abstract}


\end{frontmatter}


\section{Introduction}
The effectiveness of vaccinations is highly dependent on a reliably high immunization rate in the populace. For example, the effort to eradicate measles in the US was slowed in the late 1980's and early 1990's due to declining immunization rates, and studies have linked sporadic outbreaks following the 2000 eradication of endemic measles to deliberately under- or un-vaccinated individuals\cite{Phadke2016}.

There are indications the recent and current outbreaks are tied only in part to low overall vaccination rates, which have been relatively stable\cite{NationalCenterforHealthStatistics2016}, but also in part to local phenomena. For example high numbers of non-medical exemptions (NMEs) from vaccination schedules have been linked to outbreaks on a local\cite{Smith2004, Atwell2013} and state level\cite{Omer2006}.

This may be part in due to the differences between NME policies in various states, as evidenced by increased morbidity of preventable diseases in states with high NME rates\cite{Omer2006}, and among children with NMEs\cite{Salmon1999, Feikin2000}.
However outbreaks of preventable diseases have also been linked to areas in which vaccine skeptics, colloquially known as `antivaxxers' have led intense campaigns\cite{Hall2017}. In addition to purely empirical data, theoretical models and computer simulations have shown that interplay between awareness and disease dynamics can alter the epidemiological threshold for a disease\cite{Granell2013, Pananos2017}.

While typical misconceptions regarding vaccines are readily debunked in the literature\cite{DeStefano2019, Hviid2019}, simply presenting antivaxxers with this tends to further entrench their position\cite{Kahan2010a}.
However, many parents who interact with health care professionals are more likely to vaccinate\cite{Smith2006}. The motives for avoiding vaccines are typically diverse\cite{Helps2019, Kata2010}, and there are calls for increased communications with those skeptical of vaccination \cite{Diekema2005} to address these.

This highlights need for a better understanding of typical narratives in the vaccination debate, and of the structures through which not only infectious disease, but also information regarding the disease and preventative measures.

\section{Results}
The present section contains the results of a few different analyses. We classified tweets containing keywords related to vaccination in the period 2013 to 2019 using a deep neural network. The materials and methods section contains details on the data and classifier. From the classified tweets we identified profiles that consistently produced content expressing opposition to, and support of, vaccination. We refer to these as anti- and pro-vaxx profiles.
We investigate the interplay between user-level vaccine sentiments various behavioural patterns such as post frequency, network/community structure, etc.

\subsection*{External links (URLs)}
We searched for external links posted in tweets containing vaccination-related keywords in the period spanning from February 2015 to May 2019. We converted each link to a `base-url', for example converting 
\url{https://www.youtube.com/watch?v=thHprH4G8k8} into \url{youtube.com}.

\begin{figure}[htb]
	\centering
	\includegraphics[width=0.9\linewidth]{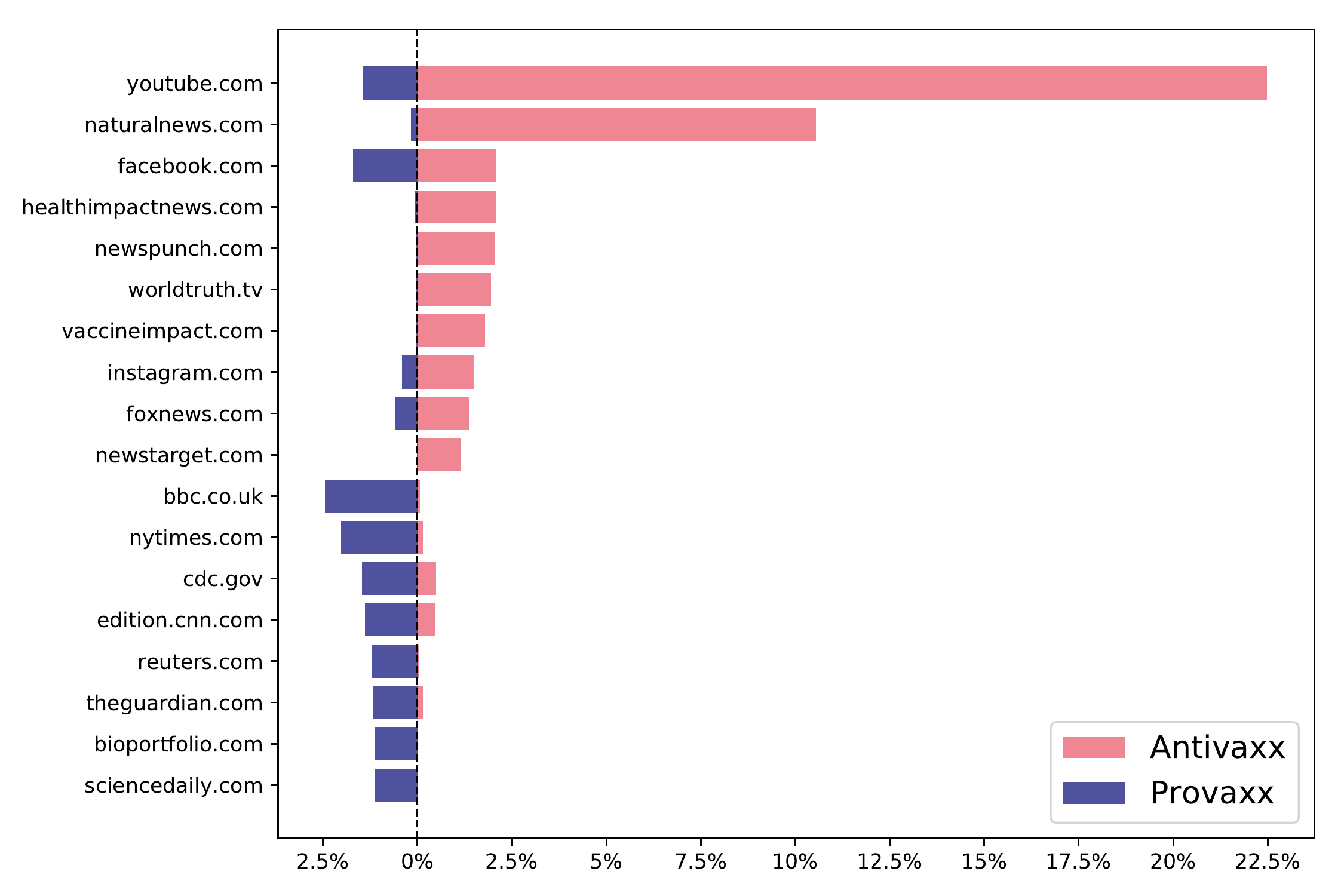}
	\caption{The top 10 most linked to domains by strongly antivaxx and provaxx profiles. Bar length shows percentage of the total number of links shared by profiles in the given category and hence do not sum to 100. For each domain, the red bars going right represent antivaxxers and blue bars going left provaxxers. Antivaxxers rely heavily on links to Youtube, and the page 'natural news', which promulgates pseudoscience and sells products related to health and nutrition. Provaxxers link to a wide array of news and science sites, which is why a lower overall percentage of their links are contained in the top 10.}
	\label{fig:links}
\end{figure}

We found the top 10 such base URLs for both anti- and pro-vaxxers. This resulted in 18 unique base URLs as Youtube and Facebook links were present in both top 10 sets. The frequency at which each of the 18 sources where linked to in both profile categories, is shown in \cref{fig:links}.

To gain an understanding of this more broadly, we extracted the top 20 links for each profile category for the entire period, an assigned categories such as 'science', 'news', etc. to each base URL. Some URLs were assigned multiple categories - for example sciencedaily.com was labeled as both 'science' and 'news'. If a products were sold on a site, it was labeled as 'commercial'. An overview of the assigned categories is shown in the Materials and Methods section.

\begin{figure}[htb]
	\centering
	\includegraphics[width=0.9\linewidth]{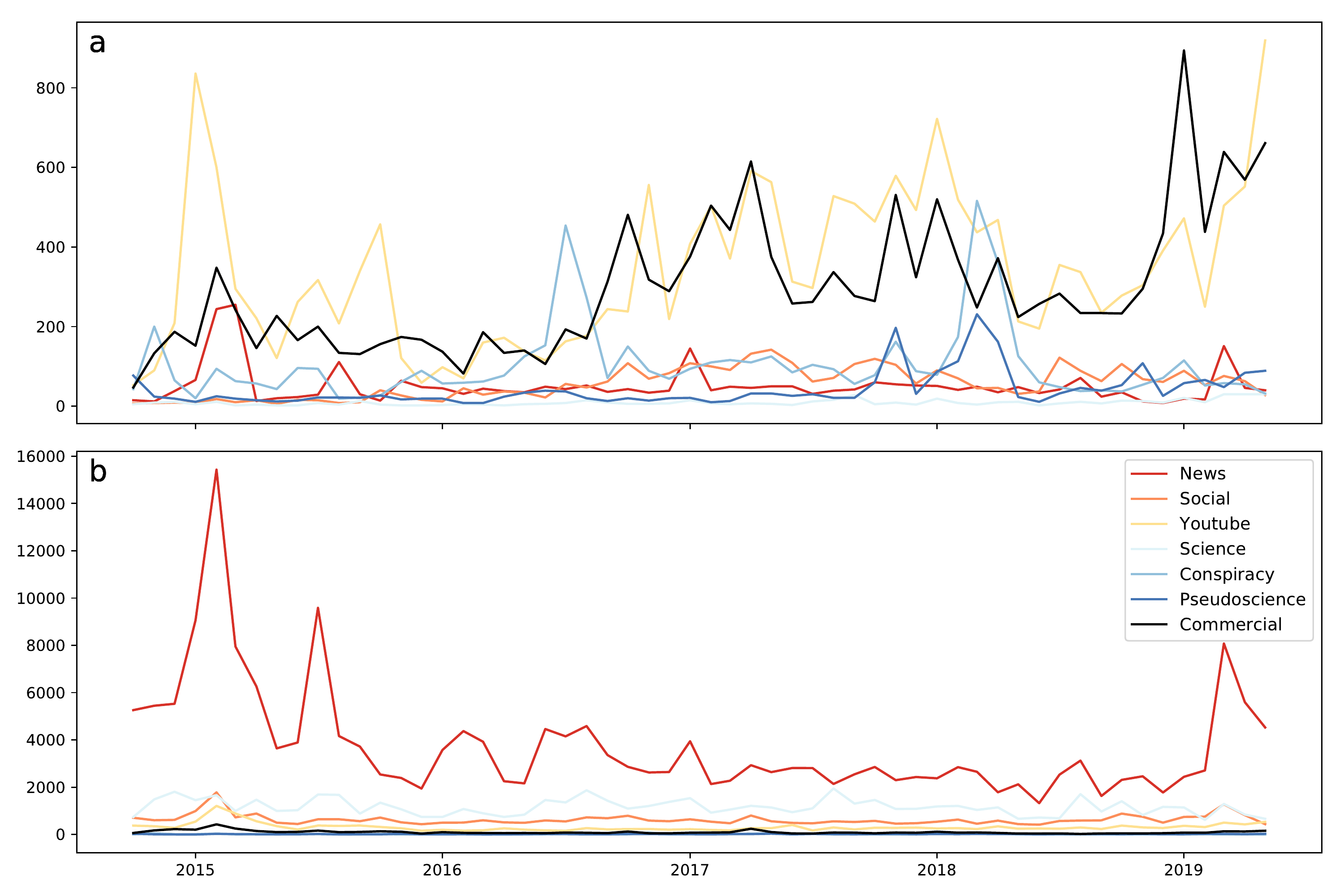}
	\caption{Frequency of various categories of links for \textbf{a} antivaxx and \textbf{b} provaxx profiles.}
	\label{fig:temporal_links}
\end{figure}

We found that, across the time period, links to Youtube and commercial sites represented the majority of the common links by antivaxxers, with conspiracy sites occasionally gaining popularity. In contrast, news sites where by far the most common for provaxxers. This is visualized in \cref{fig:temporal_links}.

\subsection*{Activity}
\begin{figure}[htb]
	\centering
	\includegraphics[width=0.9\linewidth]{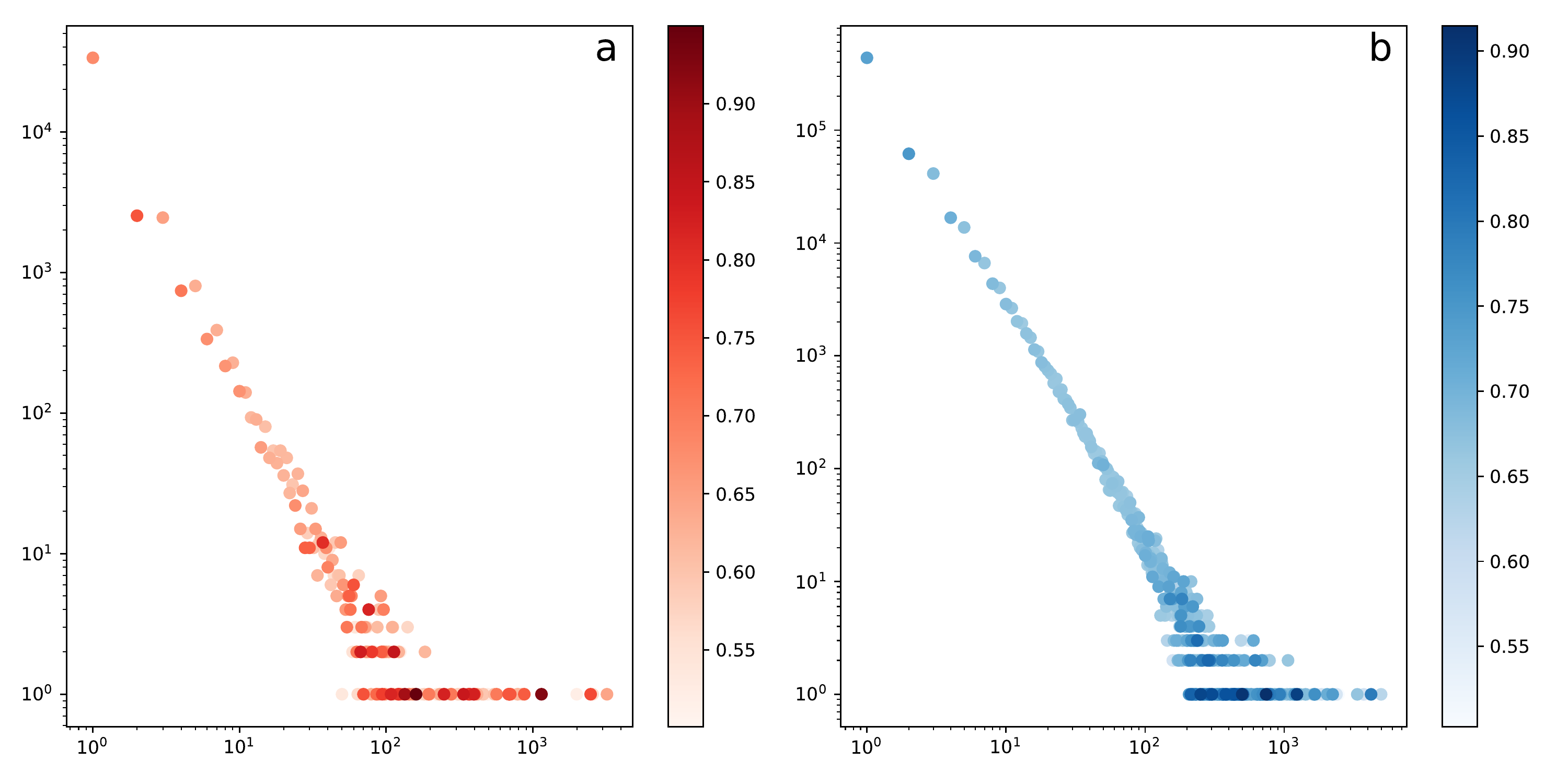}
	\caption{Interplay between vaccine sentiment and profile activity. The distribution of profile activities appears to follow a power law for both \textbf{a} anti- and \textbf{b} provaxx profiles. Color denotes the median anti/provaxx probability of vaccination-related tweets profiles with the given number of vaccine-related tweets. Profiles at the more active end of the spectrum seem to more commonly express extreme sentiment, but this might be due to the higher variance caused by dots representing fewer users.}
	\label{fig:activities}
\end{figure}

Analyzing the frequency of contributions from anti- and provaxx profiles, we found a power law-like distribution for both groups, see \cref{fig:activities}, with more active profile being responsible for a disproportionately large quantity of tweets.

\begin{figure}[htb]
	\centering
	\includegraphics[width=0.9\linewidth]{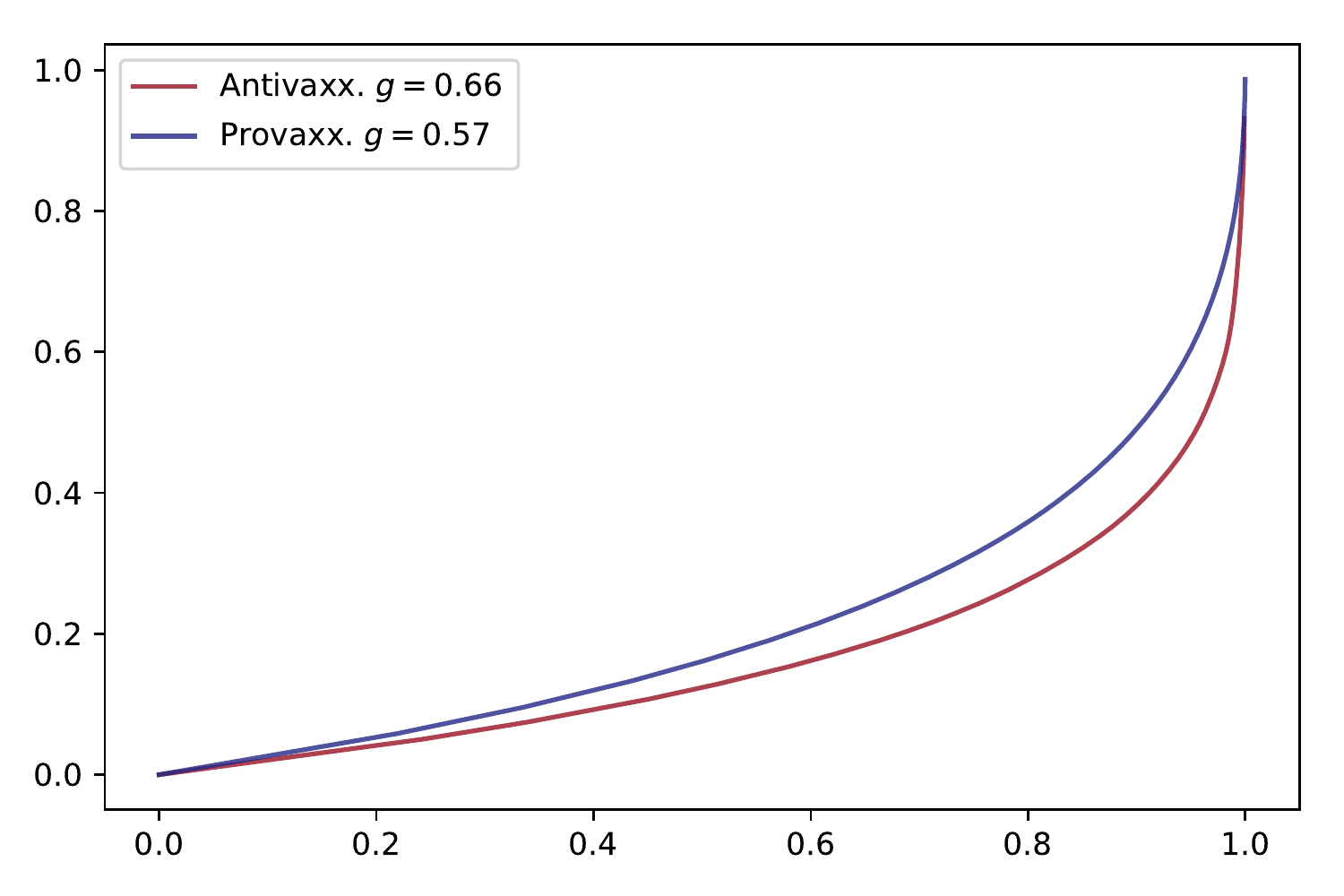}
	\caption{Lorenz curves and Gini coefficients $g$ for the two groups of profiles. The contributions of antivaxx users are more skewed in the sense that high-activity profiles contribute a more disproportionate amount of the content generated by antivaxx profiles.}
	\label{fig:vaxxgini}
\end{figure}

This effect is more pronounced for antivaxxers, where $3.75\%$ of profiles generate $50\%$ of the content, compared with $9.16\%$ for provaxxers. \Cref{fig:vaxxgini} compares activity skewness for the two groups with Lorenz curves.

In order to analyze the dynamics of the vaccination debate, we first construct a network of reciprocal interactions on Twitter in the time period from late 2013 to late 2016 (details in methods section). We refer to this network as the MMR graph, for \textit{mutual mentions/retweets}, as nodes in the graph are connected only if the users represented by the nodes have both mentioned or retweeted each other. The MMR graph exhibits a stretched exponential degree distribution, and edges show a temporal self-similarity that gradually diminishes with time, see \cref{fig:twitter}.

\subsection*{Network-based Analysis}
To investigate possible network phenomena, constructed a graph representing mutual interactions from a dataset containing a random $10\%$ sample of tweets from late 2013 to late 2016\cite{Liu2014}. Nodes in the graph represent Twitter profiles, and nodes are connected only if both profiles have interacted with the other by mentioning, replying to, or retweeting them. Therefore, we refer to this as the \textit{mutual mentions and reply/retweet graph}, or simply the \textit{MMR graph}.

\begin{figure}[htb]
	\centering
	\includegraphics[width=0.9\linewidth]{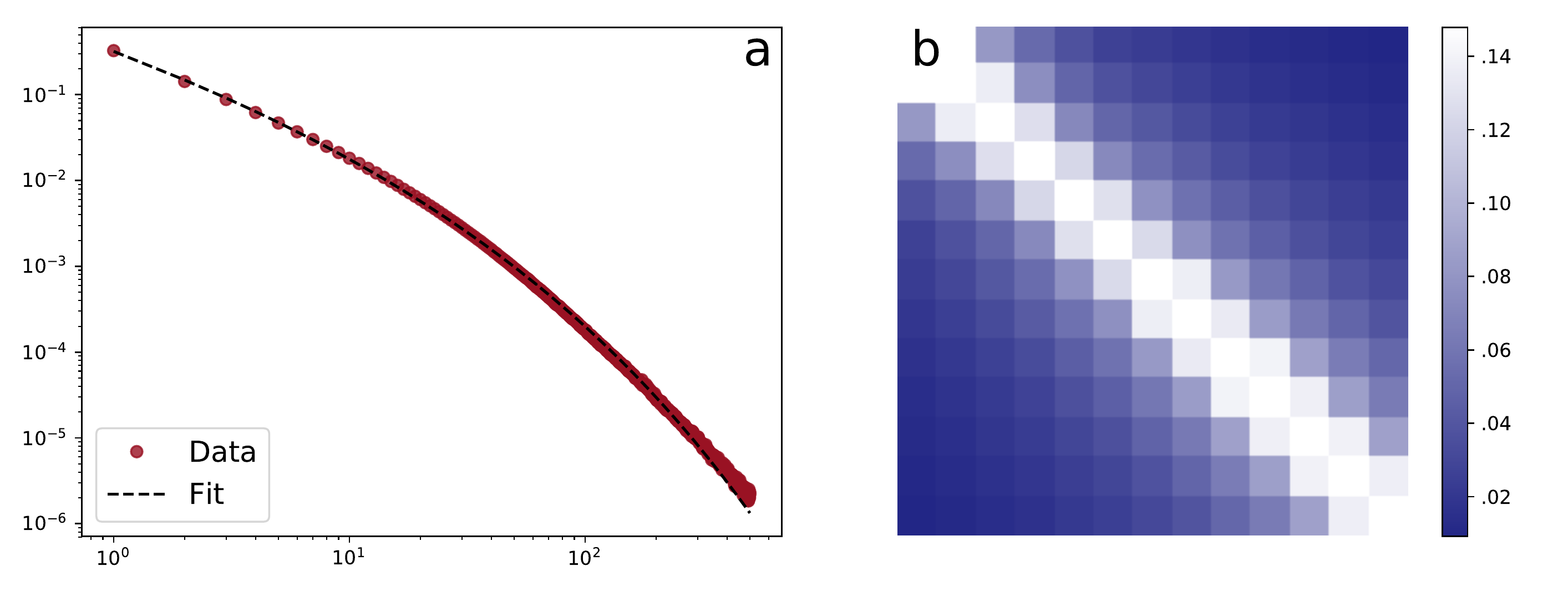}
	\caption{\textbf{a:} The degree distribution of the MMR graph, truncated at degree $500$ to exclude automated profiles. The dashed line indicates the best fit for a stretched exponential function.  \textbf{b:} The Jaccard similarity index of the sets of edges in the MMR graph for different 3-month periods.}
	\label{fig:twitter}
\end{figure}

We constructed MMR graphs for timeslices of 3 months, starting with September through November of 2013 and ending with the same months in 2016. Comparing the graph at different timeslices, it exhibits a self-similarity which decreases in time to almost nothing after 3 years, shown in \cref{fig:twitter}. Then degree distribution of the graph aggregated over all time windows is a stretched exponential (Weibull) distribution, also shown in \cref{fig:twitter}.

We then generated a subgraph of the MMR graph representing representing repeated mutual interaction of profiels expressing strong sentiments regarding vaccines. We did this by aggregating the graph, and keeping only nodes labelled as anti/provaxx, and links that occurred in at least 2 time windows, and then keeping only the giant connected component of the resulting graph.

The resulting graph contains 3359 nodes, of which 395 ($11.76\%$) represent antivaxxers. The graph is heavily stratified, with an assortativity coefficient of $r = 0.813$.

\begin{figure}[htb]
	\centering
	\includegraphics[width=0.99\linewidth]{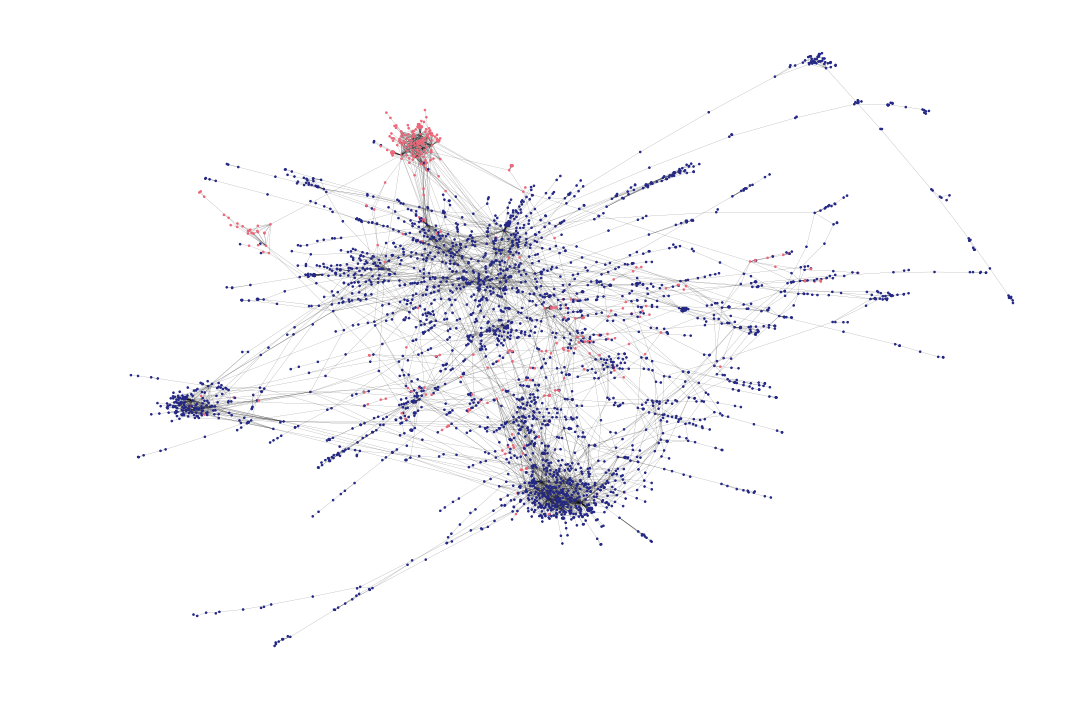}
	\caption{Representation of the repeated mutual interaction graph from 2013-2016.}
	\label{fig:polarized}
\end{figure}

We generated a layout for the graph using the force layout algorithm\cite{Jacomy2014}, the result of which is shown in \cref{fig:polarized}. The high degree of homophily is clearly visible in the graph.

\FloatBarrier
\subsection*{Materials and methods.}
This section provides details of the data analyzed and the methods employed.

\subsection*{Twitter data}
Our initial Twitter dataset consisted of a large collection (approximately 60 billion) of tweets, collected by Alan Mislove of Northeastern University\cite{Helps2019}. To collect additional data, we set up a stream listener using the tweepy module in Python. This listened for the following vaccination-related keywords: "unvaccinated",
"unvaccined",
"vaccinate",
"vaccinated",
"vaccinating",
"vaccination",
"vaccinations",
"vaccinator",
"vaccinators",
"vaccine",
"vaccined",
"vaccinering",
"vaccines",
"vaccinology",
"vaxx".

In addition, we used Python to paginate backwards through the official search API for tweets matching either of the keywords. When a match occurred, the tweet was analyzed and saved to a database. The analysis involved evaluating the sentiment expressed by the tweet (details in the subsequent section), location analysis, and link following.

Location analysis followed a three-tiered approach:
\begin{itemize}
	\item If the tweet's metadata contained raw coordinates (lattitude and longitude), those where used to directly obtain the location.
	\item If not, we extracted the 'location string' form the tweet aand tried two approaches, as follows:
	\subitem We first passed the location string to an open\-street\-map based geo\-co\-der in Python.
	\subitem If this failed, we compared the string with a repository obtained from the aforementioned Twitter data. From those data, we had a mapping from such strings unto a maximum likelihood estimation of the probability of locations in terms of US county (FIPS) codes. From that we obtained a probability distribution over state codes and saved that.
\end{itemize}

Many links posted to Twitter are shortened by external URL shorteners. To remedy this, we used Python to recursively follow redirects of each link until it resolved, timed out, or a loop was detected. As this process easily becomes input/output bottlenecked (link following took much longer than the typical time between two tweets being posted), we used a parallel approach where tweets were sent to a multiprocess pool as they were received, allowing us to crawl many links in parallel.

The links we analysed were categorized as shown in the following:
\begin{sloppypar}
\begin{itemize}
\item \textbf{commercial} \nolinkurl{articles.mercola.com}, \nolinkurl{go.thetruthaboutvaccines.com}, \nolinkurl{greenmedinfo.com}, \nolinkurl{healthimpactnews.com}, \nolinkurl{healthnutnews.com}, \nolinkurl{infowars.com}, \nolinkurl{naturalnews.com}, \nolinkurl{newstarget.com}, \nolinkurl{vaccineimpact.com}
\item \textbf{conspiracy} \nolinkurl{awarenessact.com}, \nolinkurl{newspunch.com}, \nolinkurl{newstarget.com}, \nolinkurl{therealstrategy.com}, \nolinkurl{worldtruth.tv}
\item \textbf{news} \nolinkurl{bbc.co.uk}, \nolinkurl{bbc.com}, \nolinkurl{bioportfolio.com}, \nolinkurl{cbc.ca}, \nolinkurl{choice.npr.org}, \nolinkurl{cnn.com}, \nolinkurl{edition.cnn.com}, \nolinkurl{forbes.com}, \nolinkurl{foxnews.com}, \nolinkurl{huffingtonpost.com}, \nolinkurl{medicalnewstoday.com}, \nolinkurl{nbcnews.com}, \nolinkurl{nytimes.com}, \nolinkurl{reuters.com}, \nolinkurl{sciencedaily.com}, \nolinkurl{statnews.com}, \nolinkurl{theguardian.com}, \nolinkurl{time.com}, \nolinkurl{whitehouse.gov}
\item \textbf{pseudoscience} \nolinkurl{collective-evolution.com}, \nolinkurl{inshapetoday.com}, \nolinkurl{realfarmacy.com}, \nolinkurl{seattleorganicrestaurants.com}
\item \textbf{science} \nolinkurl{bioportfolio.com}, \nolinkurl{cdc.gov}, \nolinkurl{medicalnewstoday.com}, \nolinkurl{sciencedaily.com}, \nolinkurl{statnews.com}, \nolinkurl{webmd.com}
\item \textbf{social} \nolinkurl{facebook.com}, \nolinkurl{instagram.com}, \nolinkurl{reddit.com}
\item \textbf{youtube} \nolinkurl{youtube.com}
\end{itemize}
\end{sloppypar}

\subsection*{Classification}
We trained a deep neural network to classify tweets as anti-vaccine, pro-vaccine, or neutral/unrelated, using a transfer learning approach. This approach consists of first training a complex model to a large source dataset, then stripping off the layer of output neurons and using the output of the next last, representation, layer in conjunction with another model to predict the desired target dataset. This sometimes increases model performance, as it allows one to 'reuse' higher-order representations of the input data learned by the original classifier.

for the target data, we hired workers on Amazon's Mechanican Turk (MTurk) platform to classify tweets as being either for, or agains human vaccination, or as undecideable or unrelated. To ensure high-quality ratings, we first manually rated 100 tweets, then hired a number of workers for a test assignment which clearly stated that top performers would receive offers for additional tasks. The payment was set to be very high compared to typical MTurk to provide incentive for good performance. We then identified top performers whose scores where most similar to our own, and launched the remaining tasks, allowing only the identified workers to participate. We hired them to label $10,000$ tweets with 3 raters pr tweet. We then kept only the tweets for which all 3 raters agreed on a label, which reduced the data set to 5358, the distribution of labels in which was 
$18.8\%$ antivaxx, $45.67\%$ provaxx, and $35.50\%$ neutral/unrelated.

As the source dataset, we chose to train the classifier to predict a number of hashtags which we presumed to be related to the sentiment prediction task. From an initial qualitative analysis of the data, and from a brief review of the literature, we noted that
\begin{itemize}
	\item Anti-vaccine narratives occasionally supposes underlying conspiracies, as represented by hashtags such as \#cdctruth, or \#cdcwhistleblower.
	\item Many tweets that that mention vaccine-related keywords are not concerned with vaccination of humans, but rather of pets. To help the classifier disambiguate, we included hashtags such as \#dog and \#cat.
	\item There is a relatively popular indie rock band called \textit{The Vaccines}. To help disambiguate, we included hashtags like \#music and \#livemusic.
\end{itemize}
The full list of included hashtags is as follows:
\#endautismnow, \#antivax, \#autism, \#autismismedical, \#cat, \#cdctruth, \#cdcwhistleblower, \#dog, \#ebola, \#flu, \#health, \#hearthiswell, \#hpv, \#immunization, \#livemusic, \#measles, \#medication, \#music, \#polio, \#sb277, \#science, \#vaccination, \#vaccine, \#vaccines, \#vaccinescauseautism, \#vaccineswork, \#vaxxed.

We then downloaded a large number of tweets ($\approx 10,670,000$ in total) of tweets containing either of those hashtags. and trained a deep neural network classifier to predict the hashtags from text. We used a random upsampling approach to achieve a balanced dataset.

The classifier consisted of an embedding layer, a spatial dropout, then a parallel sequence of a) a bi-directional GRU (gated recurrent unit) and a dropout layer, and b) a weighted attention average layer\cite{Felbo2017}. Those were then concatenated into a representation layer.

After fitting the hashtag model, we removed the output layer and 'froze' the remaining layers, to prohibit training of the weights contained in the original model. We then added a fasttext network\cite{Joulin2016} in parallel with the pretrained classifier. The rationale for this was that, while the initial classifier might have learned to recognize highly complex patterns in text, it might not do a good job of making simpler connections between input text and target probabilities. After fitting the fasttext part of the classifier, we used the chain-thaw approch of\cite{Felbo2017} to further improve performance.

\begin{figure}[htb]
	\centering
	\includegraphics[width=0.99\linewidth]{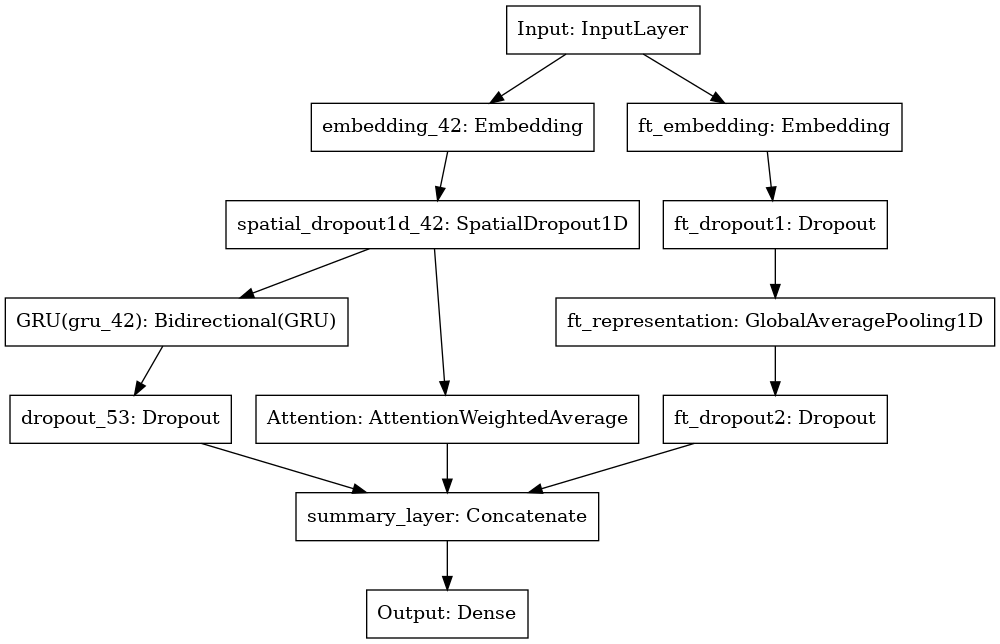}
	\caption{Representation of the final classifier.}
	\label{fig:model}
\end{figure}

On the three-class prediction task, the classifier attained a micro-averaged F1-score of $0.762$. The score was computed by aggregating true and false positives/negatives over a 10-fold stratified cross-validation procedure\cite{Forman2010}.
For comparison with the literature, we also trained the classifier for binary prediction (i.e. predicting simply whether a text snippet was anti-vaxx or not). The accuracy on the binary case was $90.4 \pm 1.4\%$ over a 10-fold stratified cross-validation evaluation, an increase over what to our knowledge is state of the art performance\cite{Mitra2016}.

Looking qualitatively at the performance of the classifier, the tweets that were labeled with high confidence demonstrate some capability of the classifier to recognize relatively subtle indications of the correct label for the tweet, as shown in \cref{tab:qualitative}.

\begin{table}
	\centering
	\begin{tabular}{p{9.5cm}c}
		\toprule
		Text & Class\\ 
		\midrule
		``January is Cervical Health Awareness Month.Join the HPV vaccine campaign to prevent cervical cancer'' & Provaxx\\
		``Getting a flu shot will help prevent transmission to new babies who are too young to be vaccinated. \#flukills'' & Provaxx\\
		``things i love about the vaccines: they change their set list every gig but norgaard is always the last song they play'' & Neutral \\ 
		``Hi $\langle \text{USER} \rangle$ £7.3m to kill 1771 Badgers | £4100 per Badger | Vaccination looks cheap now? £662 per Badger?'' & Neutral \\
		``Bill Gates Admits \#Vaccines Are Used for Human \#Depopulation'' & Antivaxx\\
		``Lead Developer Of HPV Vaccines Comes Clean, Warns Parents \& Young Girls It’s All A Giant Deadly Scam'' & Antivaxx  \\
		\bottomrule
	\end{tabular}
	\vspace*{2mm}
	\caption{Qualitative summary of classifier performance. The classifier correctly assigns a large probability of antivaxxness to text snippets the express conspiracist notions about vaccines being part of a global scam. Similarly, texts highlighting the positive qualities of vaccinations are assigned a high probability of being provaxx. In addition, text snippets concerning the band named The Vaccines are recognized as irrelevant. A text snippet expressing how much more expensive it is to kill, rather than vaccinate, badgers is also categorized as irrelevant with a high certainty, despite containing negative words like `kill'.}
	\label{tab:qualitative}
\end{table}

For each user, we counted the number of tweets concerning vaccination (based on keyword matching), and computed the probability of each tweet being characterized as pro/antivaxx, according to our classifier. We then label profiles as anti/pro-vaxx if the classifier assigns more than $50\%$ of the profile's tweets a probability of at least $50\%$ of being anti/pro-vaxx. This strong criterion is intended to reduce the number of profiles falsely assigned into either category.

\section{Acknowledgements}
The authors wish to thank Alan Mislove for his invaluable help with collection and analysis of Twitter data, and Bjarke Felbo for sharing his wisdom of machine learning.

\section{References}

\bibliography{library2}

\begin{thebibliography}{23}
\expandafter\ifx\csname natexlab\endcsname\relax\def\natexlab#1{#1}\fi
\providecommand{\url}[1]{\texttt{#1}}
\providecommand{\href}[2]{#2}
\providecommand{\path}[1]{#1}
\providecommand{\DOIprefix}{doi:}
\providecommand{\ArXivprefix}{arXiv:}
\providecommand{\URLprefix}{URL: }
\providecommand{\Pubmedprefix}{pmid:}
\providecommand{\doi}[1]{\href{http://dx.doi.org/#1}{\path{#1}}}
\providecommand{\Pubmed}[1]{\href{pmid:#1}{\path{#1}}}
\providecommand{\bibinfo}[2]{#2}
\ifx\xfnm\relax \def\xfnm[#1]{\unskip,\space#1}\fi
\bibitem[{Atwell et~al.(2013)Atwell, {Van Otterloo}, Zipprich, Winter,
  Harriman, Salmon, Halsey \& Omer}]{Atwell2013}
\bibinfo{author}{Atwell, J.~E.}, \bibinfo{author}{{Van Otterloo}, J.},
  \bibinfo{author}{Zipprich, J.}, \bibinfo{author}{Winter, K.},
  \bibinfo{author}{Harriman, K.}, \bibinfo{author}{Salmon, D.~a.},
  \bibinfo{author}{Halsey, N.~a.}, \& \bibinfo{author}{Omer, S.~B.}
  (\bibinfo{year}{2013}).
\newblock \bibinfo{title}{{Nonmedical vaccine exemptions and pertussis in
  California, 2010}}.
\newblock {\it \bibinfo{journal}{Pediatrics}\/},  {\it
  \bibinfo{volume}{132}\/}, \bibinfo{pages}{624--30}.
  \DOIprefix\doi{10.1542/peds.2013-0878}.
\bibitem[{DeStefano et~al.(2019)DeStefano, Bodenstab \& Offit}]{DeStefano2019}
\bibinfo{author}{DeStefano, F.}, \bibinfo{author}{Bodenstab, H.~M.}, \&
  \bibinfo{author}{Offit, P.~A.} (\bibinfo{year}{2019}).
\newblock \bibinfo{title}{{Principal Controversies in Vaccine Safety in the
  United States}}.
\newblock {\it \bibinfo{journal}{Clinical Infectious Diseases}\/}, .
  \DOIprefix\doi{10.1093/cid/ciz135}.
\bibitem[{Diekema(2005)}]{Diekema2005}
\bibinfo{author}{Diekema, D.} (\bibinfo{year}{2005}).
\newblock \bibinfo{title}{{Responding to parental refusals of immunization of
  children}}.
\newblock {\it \bibinfo{journal}{Pediatrics}\/}, .
\bibitem[{Feikin et~al.(2000)Feikin, Lezotte, Hamman \& Salmon}]{Feikin2000}
\bibinfo{author}{Feikin, D.}, \bibinfo{author}{Lezotte, D.},
  \bibinfo{author}{Hamman, R.}, \& \bibinfo{author}{Salmon, D.}
  (\bibinfo{year}{2000}).
\newblock \bibinfo{title}{{Individual and community risks of measles and
  pertussis associated with personal exemptions to immunization}}.
\newblock {\it \bibinfo{journal}{Jama}\/}, .
\bibitem[{Felbo et~al.(2017)Felbo, Mislove, S{\o}gaard, Rahwan \&
  Lehmann}]{Felbo2017}
\bibinfo{author}{Felbo, B.}, \bibinfo{author}{Mislove, A.},
  \bibinfo{author}{S{\o}gaard, A.}, \bibinfo{author}{Rahwan, I.}, \&
  \bibinfo{author}{Lehmann, S.} (\bibinfo{year}{2017}).
\newblock \bibinfo{title}{{Using millions of emoji occurrences to learn
  any-domain representations for detecting sentiment, emotion and sarcasm}}.
\newblock {\it \bibinfo{journal}{arXiv preprint arXiv:1708.00524}\/}, .
  \href{http://arxiv.org/abs/1708.00524}{\tt arXiv:1708.00524}.
\bibitem[{Forman \& Scholz(2010)}]{Forman2010}
\bibinfo{author}{Forman, G.}, \& \bibinfo{author}{Scholz, M.}
  (\bibinfo{year}{2010}).
\newblock \bibinfo{title}{{Apples-to-apples in cross-validation studies}}.
\newblock {\it \bibinfo{journal}{ACM SIGKDD Explorations Newsletter}\/},  {\it
  \bibinfo{volume}{12}\/}, \bibinfo{pages}{49}.
  \DOIprefix\doi{10.1145/1882471.1882479}.
\bibitem[{Granell et~al.(2013)Granell, G{\'{o}}mez \& Arenas}]{Granell2013}
\bibinfo{author}{Granell, C.}, \bibinfo{author}{G{\'{o}}mez, S.}, \&
  \bibinfo{author}{Arenas, A.} (\bibinfo{year}{2013}).
\newblock \bibinfo{title}{{Dynamical Interplay between Awareness and Epidemic
  Spreading in Multiplex Networks}}.
\newblock {\it \bibinfo{journal}{Physical Review Letters}\/},  {\it
  \bibinfo{volume}{111}\/}, \bibinfo{pages}{128701}.
  \DOIprefix\doi{10.1103/PhysRevLett.111.128701}.
\bibitem[{Hall et~al.(2017)Hall, Banerjee, Kenyon, Strain, Griffith,
  Como-Sabetti, Heath, Bahta, Martin, McMahon, Johnson, Roddy, Dunn \&
  Ehresmann}]{Hall2017}
\bibinfo{author}{Hall, V.}, \bibinfo{author}{Banerjee, E.},
  \bibinfo{author}{Kenyon, C.}, \bibinfo{author}{Strain, A.},
  \bibinfo{author}{Griffith, J.}, \bibinfo{author}{Como-Sabetti, K.},
  \bibinfo{author}{Heath, J.}, \bibinfo{author}{Bahta, L.},
  \bibinfo{author}{Martin, K.}, \bibinfo{author}{McMahon, M.},
  \bibinfo{author}{Johnson, D.}, \bibinfo{author}{Roddy, M.},
  \bibinfo{author}{Dunn, D.}, \& \bibinfo{author}{Ehresmann, K.}
  (\bibinfo{year}{2017}).
\newblock \bibinfo{title}{{Measles Outbreak — Minnesota April–May 2017}}.
\newblock {\it \bibinfo{journal}{MMWR. Morbidity and Mortality Weekly
  Report}\/},  {\it \bibinfo{volume}{66}\/}, \bibinfo{pages}{713}.
  \DOIprefix\doi{10.15585/MMWR.MM6627A1}.
\bibitem[{Helps et~al.(2019)Helps, Leask, Barclay \& Carter}]{Helps2019}
\bibinfo{author}{Helps, C.}, \bibinfo{author}{Leask, J.},
  \bibinfo{author}{Barclay, L.}, \& \bibinfo{author}{Carter, S.}
  (\bibinfo{year}{2019}).
\newblock \bibinfo{title}{{Understanding non-vaccinating parents' views to
  inform and improve clinical encounters: a qualitative study in an Australian
  community}}.
\newblock {\it \bibinfo{journal}{BMJ Open}\/},  {\it \bibinfo{volume}{9}\/},
  \bibinfo{pages}{e026299}. \DOIprefix\doi{10.1136/bmjopen-2018-026299}.
\bibitem[{Hviid et~al.(2019)Hviid, Hansen, Frisch \& Melbye}]{Hviid2019}
\bibinfo{author}{Hviid, A.}, \bibinfo{author}{Hansen, J.~V.},
  \bibinfo{author}{Frisch, M.}, \& \bibinfo{author}{Melbye, M.}
  (\bibinfo{year}{2019}).
\newblock \bibinfo{title}{{Measles, Mumps, Rubella Vaccination and Autism}}.
\newblock {\it \bibinfo{journal}{Annals of Internal Medicine}\/}, .
  \DOIprefix\doi{10.7326/M18-2101}.
\bibitem[{Jacomy et~al.(2014)Jacomy, Venturini, Heymann \&
  Bastian}]{Jacomy2014}
\bibinfo{author}{Jacomy, M.}, \bibinfo{author}{Venturini, T.},
  \bibinfo{author}{Heymann, S.}, \& \bibinfo{author}{Bastian, M.}
  (\bibinfo{year}{2014}).
\newblock \bibinfo{title}{{ForceAtlas2, a continuous graph layout algorithm for
  handy network visualization designed for the Gephi software}}.
\newblock {\it \bibinfo{journal}{PloS one}\/},  {\it \bibinfo{volume}{9}\/},
  \bibinfo{pages}{e98679}.
\bibitem[{Joulin et~al.(2016)Joulin, Grave, Bojanowski \& Mikolov}]{Joulin2016}
\bibinfo{author}{Joulin, A.}, \bibinfo{author}{Grave, E.},
  \bibinfo{author}{Bojanowski, P.}, \& \bibinfo{author}{Mikolov, T.}
  (\bibinfo{year}{2016}).
\newblock \bibinfo{title}{{Bag of tricks for efficient text classification}}.
\newblock {\it \bibinfo{journal}{arXiv preprint arXiv:1607.01759}\/}, .
\bibitem[{Kahan et~al.(2010)Kahan, Braman, Cohen, Gastil \&
  Slovic}]{Kahan2010a}
\bibinfo{author}{Kahan, D.~M.}, \bibinfo{author}{Braman, D.},
  \bibinfo{author}{Cohen, G.~L.}, \bibinfo{author}{Gastil, J.}, \&
  \bibinfo{author}{Slovic, P.} (\bibinfo{year}{2010}).
\newblock \bibinfo{title}{{Who fears the HPV vaccine, who doesn't, and why? An
  experimental study of the mechanisms of cultural cognition.}}
\newblock {\it \bibinfo{journal}{Law and Human Behavior}\/},  {\it
  \bibinfo{volume}{34}\/}, \bibinfo{pages}{501--516}.
  \DOIprefix\doi{10.1007/s10979-009-9201-0}.
\bibitem[{Kata(2010)}]{Kata2010}
\bibinfo{author}{Kata, A.} (\bibinfo{year}{2010}).
\newblock \bibinfo{title}{{A postmodern Pandora's box: anti-vaccination
  misinformation on the Internet}}.
\newblock {\it \bibinfo{journal}{Vaccine}\/}, .
\bibitem[{Liu et~al.(2014)Liu, Kliman-Silver \& Mislove}]{Liu2014}
\bibinfo{author}{Liu, Y.}, \bibinfo{author}{Kliman-Silver, C.}, \&
  \bibinfo{author}{Mislove, A.} (\bibinfo{year}{2014}).
\newblock \bibinfo{title}{{The Tweets They Are a-Changin: Evolution of Twitter
  Users and Behavior.}}
\newblock {\it \bibinfo{journal}{ICWSM}\/}, .
\bibitem[{Mitra et~al.(2016)Mitra, Counts \& Pennebaker}]{Mitra2016}
\bibinfo{author}{Mitra, T.}, \bibinfo{author}{Counts, S.}, \&
  \bibinfo{author}{Pennebaker, J.} (\bibinfo{year}{2016}).
\newblock \bibinfo{title}{{Understanding Anti-Vaccination Attitudes in Social
  Media.}}
\newblock {\it \bibinfo{journal}{ICWSM}\/}, .
\bibitem[{{National Center for Health
  Statistics}(2016)}]{NationalCenterforHealthStatistics2016}
\bibinfo{author}{{National Center for Health Statistics}}
  (\bibinfo{year}{2016}).
\newblock {\it \bibinfo{title}{{Health, United States, 2015: with special
  feature on racial and ethnic health disparities}}\/}.
\newblock \bibinfo{type}{Technical Report} National Institutes of Health.
\bibitem[{Omer et~al.(2006)Omer, Pan, Halsey, Stokley \& Moulton}]{Omer2006}
\bibinfo{author}{Omer, S.}, \bibinfo{author}{Pan, W.}, \bibinfo{author}{Halsey,
  N.}, \bibinfo{author}{Stokley, S.}, \& \bibinfo{author}{Moulton, L.}
  (\bibinfo{year}{2006}).
\newblock \bibinfo{title}{{Nonmedical exemptions to school immunization
  requirements: secular trends and association of state policies with pertussis
  incidence}}.
\newblock {\it \bibinfo{journal}{Jama}\/}, .
\bibitem[{Pananos et~al.(2017)Pananos, Bury, Wang, Schonfeld, Mohanty, Nyhan,
  Salath{\'{e}} \& Bauch}]{Pananos2017}
\bibinfo{author}{Pananos, A.~D.}, \bibinfo{author}{Bury, T.~M.},
  \bibinfo{author}{Wang, C.}, \bibinfo{author}{Schonfeld, J.},
  \bibinfo{author}{Mohanty, S.~P.}, \bibinfo{author}{Nyhan, B.},
  \bibinfo{author}{Salath{\'{e}}, M.}, \& \bibinfo{author}{Bauch, C.~T.}
  (\bibinfo{year}{2017}).
\newblock \bibinfo{title}{{Critical dynamics in population vaccinating
  behavior.}}
\newblock {\it \bibinfo{journal}{Proceedings of the National Academy of
  Sciences of the United States of America}\/},  {\it \bibinfo{volume}{114}\/},
  \bibinfo{pages}{13762--13767}. \DOIprefix\doi{10.1073/pnas.1704093114}.
\bibitem[{Phadke et~al.(2016)Phadke, Bednarczyk, Salmon \& Omer}]{Phadke2016}
\bibinfo{author}{Phadke, V.~K.}, \bibinfo{author}{Bednarczyk, R.~A.},
  \bibinfo{author}{Salmon, D.~A.}, \& \bibinfo{author}{Omer, S.~B.}
  (\bibinfo{year}{2016}).
\newblock \bibinfo{title}{{Association Between Vaccine Refusal and
  Vaccine-Preventable Diseases in the United States}}.
\newblock {\it \bibinfo{journal}{JAMA}\/},  {\it \bibinfo{volume}{315}\/},
  \bibinfo{pages}{1149}. \DOIprefix\doi{10.1001/jama.2016.1353}.
\bibitem[{Salmon et~al.(1999)Salmon, Haber, Gangarosa \& Phillips}]{Salmon1999}
\bibinfo{author}{Salmon, D.}, \bibinfo{author}{Haber, M.},
  \bibinfo{author}{Gangarosa, E.}, \& \bibinfo{author}{Phillips, L.}
  (\bibinfo{year}{1999}).
\newblock \bibinfo{title}{{Health consequences of religious and philosophical
  exemptions from immunization laws: individual and societal risk of measles}}.
\newblock {\it \bibinfo{journal}{Jama}\/}, .
\bibitem[{Smith et~al.(2004)Smith, Chu \& Barker}]{Smith2004}
\bibinfo{author}{Smith, P.}, \bibinfo{author}{Chu, S.}, \&
  \bibinfo{author}{Barker, L.} (\bibinfo{year}{2004}).
\newblock \bibinfo{title}{{Children who have received no vaccines: who are they
  and where do they live?}}
\newblock {\it \bibinfo{journal}{Pediatrics}\/}, .
\bibitem[{Smith et~al.(2006)Smith, Kennedy, Wooten \& Gust}]{Smith2006}
\bibinfo{author}{Smith, P.}, \bibinfo{author}{Kennedy, A.},
  \bibinfo{author}{Wooten, K.}, \& \bibinfo{author}{Gust, D.}
  (\bibinfo{year}{2006}).
\newblock \bibinfo{title}{{Association between health care providers' influence
  on parents who have concerns about vaccine safety and vaccination coverage}}.
\newblock {\it \bibinfo{journal}{Pediatrics}\/}, .

\end{thebibliography}

\end{document}